\documentclass[10pt,twocolumn,conference]{IEEEtran}
\usepackage{amsmath,amssymb}
\usepackage[dvips]{graphicx}
\usepackage{amsfonts}
\usepackage[mathscr]{eucal}
\usepackage{latexsym}
\usepackage{amsthm}
\usepackage{exscale}
\usepackage[mathscr]{eucal}
\usepackage{bm}
\usepackage[dvipsnames]{color}
\usepackage{cases}
\usepackage{color}
\usepackage{epsfig}
\usepackage{algorithm}
\usepackage{algorithmic}
\usepackage[verbose,nospace,sort]{cite}
\usepackage{tabularx}
\usepackage{multirow}
\usepackage{multicol}
\usepackage{balance}
\usepackage{caption}
\usepackage{subcaption}
\usepackage{setspace}

\scrollmode

\theoremstyle{definition}

\newcommand{\thr}{\mathrm{th}}
\newcommand{\disc}{\mathrm{disc}}

\usepackage{amsmath}

\graphicspath{{./figs/}}

\begin{document}
\title{Path Design for Cellular-Connected UAV with Reinforcement Learning}
\vspace{-2ex}
\author{Yong Zeng and Xiaoli Xu \\
School of Electrical and Information Engineering, The University of Sydney\\
Email: \{yong.zeng; xiaoli.xu\}@sydney.edu.au
\vspace{-4ex}}

\maketitle

\begin{abstract}
This paper studies the path design  problem for cellular-connected unmanned aerial vehicle (UAV), which aims to minimize its mission completion time while maintaining good connectivity with the cellular network. We first argue that the conventional path design approach via formulating and solving optimization problems faces several practical challenges,  and then propose a new reinforcement learning-based UAV path design algorithm by applying \emph{temporal-difference} method to directly learn the \emph{state-value function} of the corresponding Markov Decision Process. The proposed algorithm is further extended by using linear function approximation with tile coding to deal with large state space. The proposed algorithms only require the raw measured or simulation-generated signal strength as the input and are suitable for both online and offline implementations. Numerical results show that the proposed path designs can successfully avoid the coverage holes of cellular networks even in the complex urban environment.
\end{abstract}


\section{Introduction}
Unmanned aerial vehicles (UAVs) are anticipated to play an important role in future mobile communication networks \cite{1095}. Two paradigms have been envisioned for the  seamless integration of UAVs into cellular networks, namely {\it UAV-assisted wireless communications} \cite{649}, where dedicated UAVs are dispatched as aerial communication platforms to enable the wireless connectivity for devices without or with insufficient infrastructure coverage,  and {\it cellular-connected UAV} \cite{941,1012,952}, where UAVs with their own missions are connected to cellular networks as aerial user equipments (UEs). In particular, by reusing the millions of cellular base stations (BSs) worldwide,  cellular-connected UAV is regarded as a cost-effective technology to unlock the full potential of numerous UAV applications. 

Despite of its promising applications, cellular-connected UAV also faces many new challenges. In particular, as cellular networks are mainly designed to serve terrestrial UEs and the existing BS antennas are typically downtilted, an ubiquitous cellular coverage in the sky has not yet been achieved by existing long-term evolution (LTE) networks. In fact, even for future 5G-and-beyond cellular networks that are upgraded/designed to  embrace the new aerial UEs, targeting for ubiquitous sky coverage, even for some moderate range of altitude, might be too ambitious to  practically realize due to technical challenges and/or economical consideration. 
Such coverage issue is exacerbated by the more severe interference suffered by aerial UEs \cite{941,1012,952}, due to the high likelihood of having strong line of sight (LoS) links with non-associated BSs.

 Fortunately, different from terrestrial UEs that usually move randomly and thus rendering ubiquitous ground coverage essential, the UAV mobility can be completely or partially controlled. This offers an additional degree of freedom to circumvent the aforementioned coverage issue, via {\it communication-aware trajectory design}--an approach that requires no or little modifications for cellular networks to serve aerial UEs. There have been some initial research efforts towards this direction. In \cite{1080}, by applying graph theory and convex optimization, the UAV trajectory is optimized to minimize the UAV travelling time while ensuring that it is always connected with at least one BS. A similar problem is studied in \cite{1008}, by allowing  certain tolerance for disconnection. However, both \cite{1080} and \cite{1008} assume the simple circular coverage area by each cell, which relies on some strong assumptions like isotropic antennas at the BSs and free-space path loss channel model. More importantly, the communication-aware UAV trajectory design based on solving optimization problems like \cite{1080} for cellular-connected UAV and other relevant works \cite{904} for UAV-assisted communications have some critical limitations. First, formulating an optimization problem requires accurate and analytically tractable end-to-end communication models, including the antenna model, channel model, and even the local propagation environmental model. Secondly, optimization-based design also requires the perfect and usually global knowledge of the modelling parameters, which is non-trivial to acquire in practice. Last but not least, even with the accurate modelling and the perfect information of all relevant parameters, most optimization problems in modern communication systems are highly non-convex and difficult to be efficiently solved.


\begin{figure}
\centering
\includegraphics[width=0.25\textwidth]{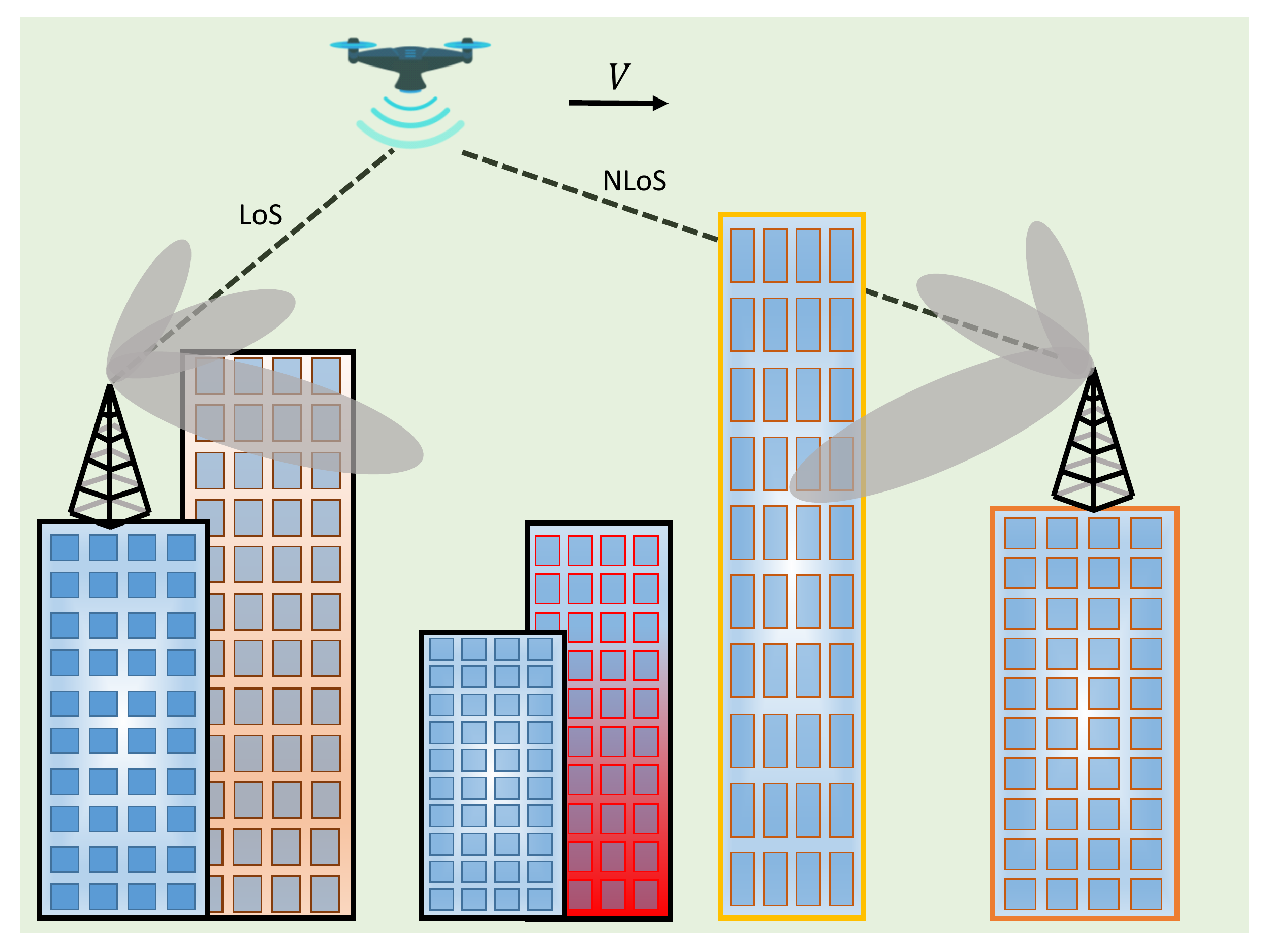}
\caption{An illustration of path design for cellular-connected UAV in urban environment.\vspace{-3ex}}
\label{F:SystemModel}
\end{figure}

To overcome the above limitations, we propose in this paper a new approach for UAV path design based on reinforcement learning (RL) \cite{1084}, which is one type of machine learning techniques for solving sequential decision problems. While RL has attracted growing attentions for wireless communications \cite{1097} in general and UAV communications in particular \cite{1057,xiaoliu2018gclearning,1096,1060}, to our best knowledge, its application to designing UAV path to avoid the cellular coverage holes (See Fig.~\ref{F:SystemModel}) has not been reported.  
 To fill the gap, we first formulate an optimization problem to minimize the weighted sum of the UAV's mission completion time and disconnection duration, and show that the formulated problem can be transformed to a Markov decision process (MDP). An efficient algorithm is then proposed for path design by applying the temporal-difference (TD) method to directly learn the state-value function of the MDP. The algorithm is further extended by using linear function approximation with tile coding so as to deal with large state space. The proposed path design algorithms can be implemented either online, offline, or a combination of both, which only require the raw measured or simulation-generated signal strength as the input.
 Numerical results show that the proposed path designs can successfully avoid the coverage holes of cellular networks even in the complex urban environment, and significantly outperform the benchmark scheme.

\section{System Model and Problem Formulation}\label{sec:SystemModel}
As shown in Fig.~\ref{F:SystemModel}, we consider a basic setup of cellular-connected UAV, which aims to design its trajectory from an initial location to a final location with a minimum flying time, while maintaining ``good'' connectivity with the cellular network. This setup corresponds to many practical UAV applications such as cellular-supported drone delivery, aerial inspection, and data collection. 
We assume that the UAV flies at a constant altitude $H$ and the horizontal coordinates of the initial and final locations are denoted by $\mathbf q_I$ and $\mathbf q_F\in \mathbb R^{2 \times 1}$, respectively. Let $T$ denote the mission completion time and $\mathbf q(t)\in \mathbb{R}^{2\times 1}$, $t\in [0, T]$, represent the UAV trajectory. We then have $\mathbf q(0)=\mathbf q_I$ and $\mathbf q(T)=\mathbf q_F$. Assume that the feasible region where the UAV can fly is a rectangular area $[x_L, x_U] \times [y_L, y_U]$. Define $\mathbf q_L=[x_L,y_L]^T$ and $\mathbf q_U=[x_U,y_U]^T$. We then have $\mathbf q_L \preceq \mathbf q(t) \preceq \mathbf q_U$, $\forall t\in [0, T]$, where $\preceq$ denotes element-wise inequality.

Let $M$ denote the number of cells that may potentially impact the UAV's path design, and $h_m(t)$ represent the end-to-end channel coefficient from cell $m$ to the UAV, which includes the transmit and receive antenna  gains, the large-scale path loss and shadowing, as well as  the small-scale fading due to multi-path propagation. As the proposed RL-based path design does not rely on any assumption on the channel modelling, the detailed discussion of one practical BS-UAV channel model is deferred to Section~\ref{sec:Numerical}. The  average received signal power by the UAV from cell $m$, with the average taken over the small scale-fading, is
\begin{align}
\bar p_m(t)=\bar P_m \mathbb E [|h_m(t)|^2], \ m=1,\cdots, M, \label{eq:pmt}
\end{align}
where $\bar P_m$ is the transmit power of cell $m$. We say that the UAV is disconnected from the cellular network at time $t$ if its received signal quality, which is  a function of the average received signal powers from the $M$ cells, is below a certain threshold $\gamma_{\thr}$, i.e., when  $f(\bar p_1,\cdots, \bar p_M)<\gamma_{\thr}$.  Two typical examples of $f(\cdot)$ is the {\it maximum received power}, where $f(\bar p_1,\cdots, \bar p_M)\triangleq \underset{m=1,\cdots, M}{\max}\bar p_m$, and the received {\it signal-to-interference ratio (SIR)},
where
$f(\bar p_1,\cdots, \bar p_M)\triangleq \frac{\bar p_{m^{\star}}}{\sum_{m\neq m^{\star}}\bar p_m}$
with $m^{\star} =\underset{m=1,\cdots, M}{\mathrm{argmax}}\ \bar p_m$.
Define an indicator function
\begin{align}\label{eq:It}
I(t)=\begin{cases}
1, \ \text{if } f\big(\bar p_1(t),\cdots, \bar p_M(t)\big)< \gamma_{\thr}\\
0, \ \text{otherwise}.
\end{cases}
\end{align}
Then the total UAV disconnection duration $T_{\disc}$ can be represented as
\begin{align}
T_{\disc}(\{\mathbf q(t)\})=\int_0^T I(t) dt.
\end{align}
It is not difficult to see that $T_{\disc}$ is a function of the UAV trajectory $\mathbf q(t)$, since the average received signal power $\bar p_m(t)$ in \eqref{eq:pmt} depends on $\mathbf q(t)$ via $h_m(t)$.

Intuitively, with larger mission completion time $T$, the UAV has higher degrees of freedom to design its trajectory to avoid the cellular coverage holes and thus reduce $T_{\disc}$. Our objective is to design  $\mathbf q(t)$ to achieve a flexible tradeoff between minimizing $T$ and $T_{\disc}$. This can be attained by minimizing the weighted sum of these two metrics with certain weight $\mu \geq 0$:
\begin{align}
\mathrm{(P0):} \underset{T, \{\mathbf q(t)\}}{\min}   \ & T + \mu T_{\disc}(\{\mathbf q(t)\}) \notag \\
\mathrm{s.t.} \  &\|\dot{\mathbf q}(t)\| \leq V_{\max}, \ \forall t\in [0, T], \\
& \mathbf q(0)=\mathbf q_I, \ \mathbf q(T)=\mathbf q_F, \label{eq:qFConstr}\\
& \mathbf q_L \preceq \mathbf q(t) \preceq \mathbf q_U, \ \forall t\in [0,T], \label{eq:RectConstr}
\end{align}
where $V_{\max}$ denotes the maximum UAV speed. It can be shown that at the optimal solution to $\mathrm{(P0)}$, the UAV should always fly with the maximum speed $V_{\max}$, i.e., we  have $\dot{\mathbf q}(t)=V_{\max} \vec{\mathbf v}(t)$, where $\vec{\mathbf v}(t)$ with $\|\vec{\mathbf v}(t)\|=1$ denotes the UAV flying direction. Thus, $\mathrm{(P0)}$ can be equivalently written as
\begin{align}
\mathrm{(P1):} \underset{T, \{\mathbf q(t), \vec{\mathbf v}(t)\}}{\min}   \ & T + \mu T_{\disc}(\{\mathbf q(t)\}) \notag \\
\mathrm{s.t.} \  &\dot{\mathbf q}(t) = V_{\max}\vec{\mathbf v}(t), \ \forall t\in [0, T], \label{eq:DifferentialEq}\\
& \|\vec{\mathbf v}(t)\|=1, \ \forall t\in [0, T], \\
& \eqref{eq:qFConstr}, \ \eqref{eq:RectConstr}. \notag
\end{align}

In practice, designing the UAV path by solving the optimization problems like   $\mathrm{(P0)}$ or $\mathrm{(P1)}$ faces several challenges, including the need to obtain an accurate and analytically tractable  expression for  $T_{\disc}(\{\mathbf q(t)\})$, the requirement of perfect information of the modelling parameters, as well as the difficulty to obtain efficient solutions due to the non-convexity of problems like $\mathrm{(P1)}$.  In the following, we propose a new approach for UAV path design by leveraging the powerful mathematical framework of RL,  which only requires the raw measured or simulation-generated signal strength as the input, without assuming any prior knowledge on the environment.




\section{Path Design with Reinforcement Learning}
\subsection{An Overview of Reinforcement Learning}\label{sec:MDPOverview}
This subsection aims to give a very brief overview on RL and settle down the key notations. 
RL is a useful machine learning framework to solve MDP \cite{1084}, which consists of an {\it agent} and the {\it environment} that  interact with each other iteratively. With fully observable MDP, at each discrete time step $n$, the agent observes a state $S_n$, takes an action $A_n$, and then receives an immediate reward $R_n$ and  transits to the next state $S_{n+1}$. Mathematically, a MDP can be specified by 4-tuple $<\mathcal S, \mathcal A, \mathcal P, \mathcal R>$, where
$\mathcal S$: the state space; $\mathcal A$: the action space; $\mathcal P$: the state transition probability, with $P(s^\prime | s, a)$ specifying the probability of transiting to the next state  $s^\prime\in \mathcal S$ given the current state $s\in \mathcal S$ after applying the action $a\in \mathcal A$; and $\mathcal R$:  the immediate reward $R(s,a)$ received by the agent.


The agent's actions are governed by its policy $\pi: \mathcal S \times \mathcal A \rightarrow [0,1]$, where $\pi(a|s)$ gives the probability of taking action $a\in \mathcal A$ when in state $s\in \mathcal S$. 
 The  goal of the agent is to improve its policy $\pi$ based on its experience, so as to maximize its long-term expected {\it return} $\mathbb{E}[G_n]$, where the return $G_n\triangleq \sum_{k=0}^{\infty} \gamma^k R_{n+k}$ is the accumulated discounted  reward from time step $n$ onwards with a discount factor $0\leq \gamma \leq 1$. 

A key notion of RL is the {\it value function}, which includes {\it state-value function} and {\it action-value function}. The state-value function of a state $s$ under policy $\pi$, denoted as $v_{\pi}(s)$, is the expected return starting from state $s$ and  following policy $\pi$ thereafter, i.e., $v_\pi(s)=\mathbb E_\pi [G_n| S_n=s]$.
Similarly, the action-value function  of taking action $a$ at  state $s$ under policy $\pi$, denoted as $q_{\pi}(s,a)$, is the expected return starting from state $s$, taking the action $a$, and following policy $\pi$ thereafter, i.e.,
$q_{\pi}(s,a)=\mathbb E_\pi [G_n| S_n=s, A_n=a]$.
The optimal state-value function, denoted as $v_*$, is defined as $v_*(s)=\underset{\pi}{\max} \ v_{\pi}(s)$, $\forall s\in \mathcal S$. Similar definition holds for the optimal action-value function. 
If the optimal value functions $q_*(s,a)$ or $v_*(s)$  is known, the optimal policy can be easily obtained either directly or with one-step-ahead search.
Thus, the essential task of many RL algorithms is to obtain the optimal value functions, which satisfy the celebrated Bellman optimality equation
\begin{align}
v_*(s)=\underset{a\in \mathcal A}{\max}\left[R(s,a)+\gamma\sum_{s^\prime \in \mathcal S} P(s'|s,a)v_*(s') \right], \ \forall s\in \mathcal S.\notag
\end{align}
Similar Bellman optimality equation holds for the action-value function. Bellman optimality equation is non-linear, where there is no closed-form solution in general. However, many iterative solutions have been proposed, such as {\it model-based dynamic programming (DP)} and {\it model-free TD learning}. 
In particular, when the agent has no prior knowledge about the environment of the MDP, it may apply the important idea of {\it TD learning}, which is a class of  model-free RL methods that learn the value functions based on the direct samples of the {\it state-action-reward-nextState} sequence, with the estimation of the value functions updated by the concept of {\it bootstrapping}.  
The simplest TD method makes the following update to the value function with an observed sample $(S_n, R_n, S_{n+1})$ \cite{1084}
\begin{align}
V(S_n)\leftarrow V(S_n) + \alpha \left[ R_n + \gamma V (S_{n+1})-V(S_n)\right],\notag
\end{align}
where $\alpha$ is the learning rate.


\subsection{UAV Path Design as an MDP}\label{sec:UAVasMDP}
The first step to apply  RL algorithms for solving a real-world problem is to formulate it as an MDP. As MDP is defined over discrete time steps, for the UAV path design problem $\mathrm{(P1)}$, we need to first discretize the time horizon $[0,T]$ into $N$ time steps with certain interval $\delta_t$. Apparently, $\delta_t$ should be sufficiently small so that within each time step, the average received signal power by the UAV in \eqref{eq:pmt} remains approximately unchanged. 
As such, the UAV trajectory $\{\mathbf q(t),\vec{\mathbf v}(t)\}$ can be specified by its discretized representation $\mathbf q[n]=\mathbf q(n\delta_t)$ and $\vec{\mathbf v}[n]=\vec{\mathbf v}(n\delta_t)$. Similarly for the average received signal power in \eqref{eq:pmt}, where $\bar p_m[n]=\bar p_m(n\delta_t)$. As a result, $\mathrm{(P1)}$ can be re-written as
\begin{align}
\mathrm{(P2):} \underset{N, \{\mathbf q[n], \vec{\mathbf v}[n]\}}{\max} \vspace{-0.1in}  \ & -(N-1) - \mu \sum_{n=0}^N I[n] \notag \\
\mathrm{s.t.} \  & \mathbf q[n+1]=\mathbf q[n]+\Delta \vec{\mathbf v}[n], \ \forall n, \label{eq:stateTrans}\\
& \|\vec{\mathbf v}[n]\|=1, \ \forall n, \\
& \mathbf q[0]=\mathbf q_I, \mathbf q[N]=\mathbf q_F,\\
& \mathbf q_L \preceq \mathbf q[n] \preceq \mathbf q_U, \forall n,
\end{align}
where \eqref{eq:stateTrans} is the discrete-time representation of the differential equation \eqref{eq:DifferentialEq} with $\Delta=V_{\max}\delta_t$,  $I[n]$ is the discrete-time counterpart of the indicator function \eqref{eq:It}, i.e., $I[n]=1$ if $f(\bar p_1[n],\cdots, \bar p_M[n])<\gamma_{\thr}$ and $I[n]=0$ otherwise. Note that we have ignored the constant factor $\delta_t$ in the objective function of $\mathrm{(P2)}$. A natural mapping of $\mathrm{(P2)}$ to an MDP $<\mathcal S, \mathcal A, \mathcal P, \mathcal R>$ thus follows:
\begin{itemize}
\item $\mathcal S$: the state space constitutes all possible UAV locations within the feasible region, i.e., $\mathcal S=\{\mathbf s: \mathbf q_L \preceq \mathbf s \preceq \mathbf q_U\}$.
\item $\mathcal A$: the action space corresponds to the UAV flying direction, i.e., $\mathcal A=\{\vec{\mathbf v}: \|\vec{\mathbf v}\|=1\}$.
\item $\mathcal P$: the state transition probability is deterministic governed by \eqref{eq:stateTrans}, or in the probabilistic form as
    \begin{align}\label{eq:StateTransProb}
    P(\mathbf s^\prime | \mathbf s, \vec{\mathbf v})=\begin{cases} 1, & \text{ if } \substack{\left(\mathbf s^\prime = \mathbf s+ \Delta \vec{\mathbf v} \text{ and } \mathbf s^\prime \in \mathcal S \right)\\ \text{ or } \left(\mathbf s^\prime = \mathbf s \text{ and } \mathbf s+ \Delta \vec{\mathbf v} \notin \mathcal S \right)} \\
    0, & \text{ otherwise.}
    \end{cases}
    \end{align}
Note that \eqref{eq:StateTransProb} ensures a feasible solution of $\mathrm{(P2)}$, since if an action $\vec{\mathbf v}$ would let the UAV out of $\mathcal S$, its location will remain unchanged.
\item $\mathcal R$: the reward $R(\mathbf s)=-1$ if the location $\mathbf s$ is covered by the cellular network and $R(\mathbf s)=-1-\mu$ otherwise.
\end{itemize}

With the above MDP formulation, it is observed that the objective function of $\mathrm{(P2)}$ corresponds to the undiscounted  (i.e., $\gamma=1$) accumulated rewards over one {\it episode} up to time step $N$, i.e., $G_0=\sum_{k=0}^N R_k$. This corresponds to one particular form of MDP, namely the {\it episodic tasks}, which are tasks containing a special state called the {\it terminal state} that separates the agent-environment interactions into {\it episodes}. 
After being formulated as a MDP, $\mathrm{(P2)}$ can be solved by applying various RL algorithms. In the following, we first apply the standard TD learning method to learn the state-value function with state-action discretization, and then extend the algorithm by using linear function approximation with tile coding.

\subsection{TD Learning with State-Action Discretization}\label{sec:TD}
Both the state and action spaces for the MDP defined in  Section~\ref{sec:UAVasMDP} are continuous. While there are various ways to directly handle continuous state-action MDP problems, the most straightforward approach is to discretize them to form a finite-state MDP. By uniformly discretizing  the action space $\mathcal A$ into $K$ values, we have $\hat {\mathcal A}=\{\vec{\mathbf v}_1,\cdots, \vec{\mathbf v}_K\}$, where $\vec{\mathbf v}_k=[\cos \phi_k, \sin \phi_k]^T$,  with $\phi_k=2\pi(k-1)/K$, $k=1,\cdots, K$. With the finite action space $\hat{\mathcal A}$ and the deterministic state-transition \eqref{eq:StateTransProb}, the corresponding discretized state space of $\mathcal S=\{\mathbf s: \mathbf q_L \preceq \mathbf s \preceq \mathbf q_U\}$ can be obtained accordingly, which is denoted as $\hat {\mathcal S}=\{\mathbf s_1, \cdots, \mathbf s_J\}$,  with $J$ representing the total number of discretized states.  With such discretizations, the UAV path design problem is quite similar to the gridworld problem \cite{1084}, but  instead of having  equal  and known rewards, the reward for the studied problem depends on whether the UAV enters a state covered by the cellular network or not.

If the UAV has the perfect knowledge of the MDP, which is the global coverage map for the considered problem,  then the standard DP algorithms such as value iteration can be applied to find the optimal UAV path. For scenarios when the UAV has no prior knowledge on the environment, we propose the model-free UAV path design algorithm based on TD learning method, which is summarized in Algorithm~\ref{Algo:TD}.

\begin{algorithm}
\caption{UAV Path Design with TD Learning.}\label{Algo:TD}
\begin{algorithmic}[1]
\STATE {\bf Initialize:} the maximum number of episodes $\bar{N}_{\mathrm{epi}}$,  maximum number of steps per episode $\bar{N}_{\mathrm{step}}$,  learning rate parameter $N_{\alpha}$, and exploration parameter $N_{\epsilon}$,
\STATE {\bf Initialize:} the state-value function $V(\mathbf s)$, $\forall \mathbf s\in \hat{\mathcal S}$.
\FOR{$n_{\mathrm{epi}}=1,\cdots, \bar{N}_{\mathrm{epi}}$}
\STATE $\alpha=\frac{N_{\alpha}}{N_{\alpha}+n_{\mathrm{epi}}}$, $\epsilon=\frac{0.5 N_{\epsilon}}{N_{\epsilon}+n_{\mathrm{epi}}}$
\STATE Initialize the state as $\mathbf s\leftarrow \mathbf q_I$, and time step $n\leftarrow 0$.
\REPEAT
\STATE Measure (or simulate) the average received signal power $\{\bar p_m\}_{m=1}^M$ at state $\mathbf s$ and let
\begin{align}
R=\begin{cases} -1-\mu, & \text{ if } f(\bar p_1,\cdots, \bar p_M)  < \gamma_{\thr} \\
-1, & \text{ otherwise}.
\end{cases}
\end{align}
\STATE Choose action $\vec{\mathbf v}$ from $\hat {\mathcal A}$ based on the $\epsilon$-greedy policy derived from $\{V(\mathbf s)\}$, i.e., $\vec{\mathbf v}=\vec{\mathbf v}_k$, where
\begin{equation}\label{eq:epsGreedy}
\small
\begin{aligned}
k=\begin{cases}
\mathrm{randi}(K), &  \text{ with prob. }  \epsilon, \\
\underset{j=1,\cdots, K}{\mathrm{argmax}} \left[R + V(\mathbf s^\prime(\mathbf s, \vec{\mathbf v}_j))\right], & \text{ with prob. } 1-\epsilon,
\end{cases}
\end{aligned}
\end{equation}
where $\mathrm{randi}(K)$ uniformly generates a random integer from $\{1,\cdots, K\}$, and $\mathbf s^\prime(\mathbf s, \vec{\mathbf v}_j)$ is the predicted next state if action $\vec{\mathbf v}_j$ is applied as governed by the  deterministic transition \eqref{eq:StateTransProb}.
\STATE Take action $\vec{\mathbf v}$ and observe the next state $\mathbf s^\prime$.
\STATE Update $V(\mathbf s)\leftarrow V(\mathbf s) + \alpha \left[ R + V(\mathbf s^\prime) -V(\mathbf s)\right]$.
\STATE Update $\mathbf s \leftarrow \mathbf s^\prime$ and $n\leftarrow n+1$.
\UNTIL $\mathbf s = \mathbf q_F$ or $n=\bar{N}_{\mathrm{step}}$.
\ENDFOR
\end{algorithmic}
\end{algorithm}

Note that in Algorithm~\ref{Algo:TD}, the TD method is applied to  learn the state-value function $V(\mathbf s)$, instead of the action-value function $Q(\mathbf s, \vec{\mathbf v})$ as in the classic {\it Q learning}. This is due to the fact that for the studied path design problem, the state-transition is deterministic and known, for which the $\epsilon$-greedy policy can be directly obtained from the state-value function via one-step-ahead search, as in \eqref{eq:epsGreedy}. This helps reduce the number of variables from $KJ$ to $J$. In Algorithm~\ref{Algo:TD}, the learning rate $\alpha$ and the exploration parameter $\epsilon$ decrease with the episode number $n_{\mathrm{epi}}$ as in Step 4, which encourages learning and exploration at early stages while promoting exploitation as $n_{\mathrm{epi}}$ gets sufficiently large.


While theoretically, the convergence of Algorithm~\ref{Algo:TD} is guaranteed for any initialization of the state-value function \cite{1084}, in practice, a random or all-zero initialization of $V(\mathbf s)$ may require infinite time steps for the UAV to reach the destination $\mathbf q_F$. Intuitively, $V(\mathbf s)$ should be initialized in a way such that in the first episode when the UAV has completely no knowledge about the radio environment, a reasonable trial should be selecting actions for the shortest path flying. Thus, we propose the {\it distance-based} value function initialization for Algorithm~\ref{Algo:TD}, with $V(\mathbf s)\leftarrow -\|\mathbf s-\mathbf q_F\|$, $\forall \mathbf s\in \hat{\mathcal S}$. 

\subsection{TD Learning with Tile Coding}\label{sec:TDTileCoding}
The TD learning method in Algorithm~\ref{Algo:TD} is known as {\it table-based}, which requires storing and updating $J$ values, each for one state, and the state value is updated only when that state is actually visited. This  becomes impractical for continuous state or when the number of discretized states $J$ is large. In order to practically apply many RL algorithms, one may resort to the useful technique of {\it function approximation} \cite{1084}, where the state-value function is approximated by certain parametric function $V(\mathbf s)\approx \hat V(\mathbf s, \boldsymbol \theta)$, $\forall \mathbf s\in \mathcal S$, with a parameter vector $\boldsymbol \theta \in \mathbb{R}^{d\times 1}$. Function approximation brings two advantages over table-based RL. Firstly, instead of storing and updating the value functions for all states, one only needs to learn the parameter $\boldsymbol \theta$, which typically has  lower dimension than the number of states, i.e., $d\ll J$. Secondly, function approximation enables {\it generalization}, i.e., the ability to predict the state-values even for those states that have never been visited, since different states are coupled with each other. 
A common metric for updating $\boldsymbol \theta$ is mean squared error (MSE), where $\mathrm{MSE}(\boldsymbol \theta)\triangleq \mathbb{E}_{\mathbf s}[V(\mathbf s)-\hat V(\mathbf s,\boldsymbol \theta)]^2$. 

The simplest function approximation is {\it linear approximation}, where $\hat V(\mathbf s, \boldsymbol \theta)\triangleq \mathbf x^T(\mathbf s) \boldsymbol \theta$, with $\mathbf x(\mathbf s)\in \mathbb{R}^{d\times 1}$ referred to as the {\it feature vector} of state $\mathbf s$. With linear function approximation, for each state-reward-nextState transition $(\mathbf s, R, \mathbf s^\prime)$ observed by the agent, $\boldsymbol \theta$ can be updated to minimize $\mathrm{MSE}(\boldsymbol \theta)$ based on the stochastic semi-gradient method \cite{1084}. For the TD method with one-step bootstrapping, we have
\begin{align}
\boldsymbol \theta \leftarrow  \boldsymbol \theta + \tilde{\alpha} \left( R+ \mathbf x^T(\mathbf s^\prime)\boldsymbol \theta-\mathbf x^T(\mathbf s)\boldsymbol \theta)\right)\mathbf x(\mathbf s), \label{eq:updateTheta}
\end{align}
where $\tilde{\alpha}$ determines the learning rate.

\begin{figure}
\centering
\includegraphics[width=0.3\textwidth]{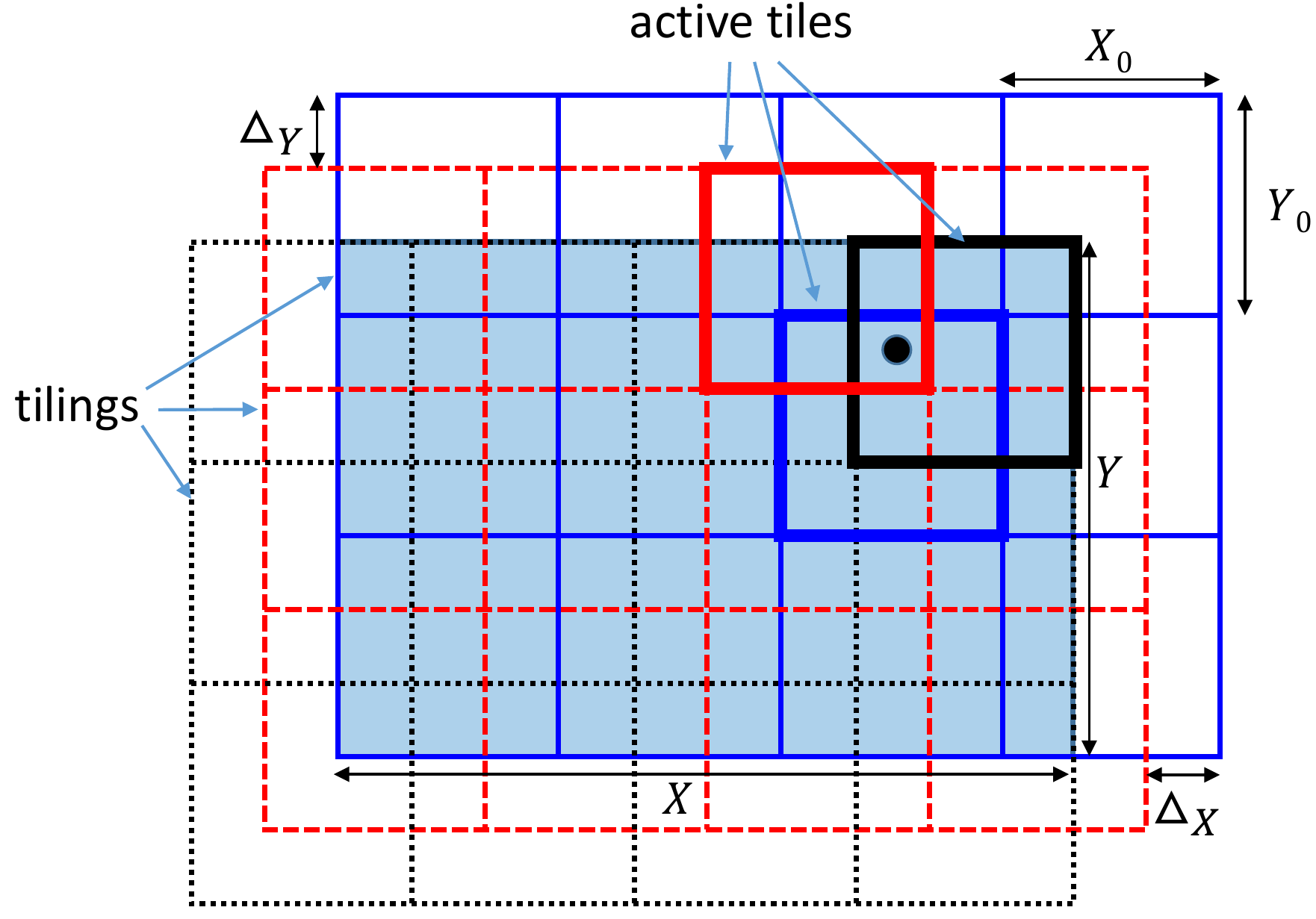}
\caption{An illustration of tile coding with 3 tilings and 12 tiles per tiling (redrawn based on Fig. 9.9 of \cite{1084}).\vspace{-3ex}}
\label{F:TildeCoding}
\end{figure}

The remaining task is to construct the feature vector $\mathbf x(\mathbf s)$. In this paper, we propose to use {\it tile coding}  \cite{1084} for feature vector construction for UAV path design. Tile coding can be regarded as a more general form of state space discretization. For the 2D rectangular area $[x_L, x_U]\times [y_L, y_U]$, instead of directly discretizing it into {\it non-overlapping} grids with sufficiently small grid size as in Section~\ref{sec:TD}, with tile coding, it is partitioned into grids with larger size, but there are many such partitions that are offset from one another by a uniform amount in each dimension. Each  such partition is called a {\it tiling} and each element of the partition is called a {\it tile}. Fig.~\ref{F:TildeCoding} gives an illustration with 3 tilings, each having 12 tiles.

As shown in Fig.~\ref{F:TildeCoding}, let $X$ and $Y$ denote the length and width of the rectangular area, respectively,  $N_1$ denote the number of tilings, and $X_0 \times Y_0$ denote the size of each tile. Then the offset between adjacent tilings can be shown to be $\Delta_X=X_0/N_{1}$ and $\Delta_Y=Y_0/N_{1}$ along the x- and y- dimensions, respectively. Let $N_2=L_XL_Y$ denote the number of tiles for each tiling. Then $L_X$ should be large enough to cover the length $X$ even after offset. Based on Fig.~\ref{F:TildeCoding}, we have $(L_X-1)X_0+\Delta_X\geq X$, or $L_X=\lceil \frac{X-\Delta_X}{X_0}\rceil +1=\lceil \frac{X}{X_0}-\frac{1}{N_1} \rceil+1$. Similar relationship can be obtained for  $L_Y$. Thus, the number of tiles for each tiling is $N_2=\left(\lceil \frac{X}{X_0}-\frac{1}{N_1} \rceil+1 \right)\left( \lceil \frac{Y}{Y_0}-\frac{1}{N_1} \rceil+1\right)$, and the total number of tiles with all tilings is $N_1N_2$. It is not difficult to see that while tiles of the same tiling are non-overlapping, those from different tilings may overlap with each other. This renders it possible to represent each point in the space by specifying the active tile of each tiling, which requires exactly $N_1$ variables. However, an effective way of representation is to use a binary vector $\mathbf x(\mathbf s)$ of dimension $N_1N_2$, with each element corresponding to one tile resulting from the $N_1$ tilings. $\mathbf x(\mathbf s)$ is a sparse vector with most elements being $0$ except for the $N_1$ elements corresponding to the active tiles in each tiling. This gives the feature vector of linear function approximation with tile coding.

The pseudo-code of TD learning with tile coding is quite similar to Algorithm~\ref{Algo:TD}, with the following straightforward modifications:
(i) Replace the state-value function  $V(\mathbf s)$ by $\hat V(\mathbf s, \boldsymbol \theta)=\mathbf x^T(\mathbf s)\boldsymbol \theta$, if $\mathbf s \neq \mathbf q_F$, and $V(\mathbf s)=0$ for $\mathbf s=\mathbf q_F$; (ii) Replace the value function update in Step 10 of Algorithm~\ref{Algo:TD} with the parameter update \eqref{eq:updateTheta}. Besides, to have the same learning rate $\alpha$ as in Algorith~\ref{Algo:TD}, the parameter $\tilde{\alpha}$ in \eqref{eq:updateTheta} should be set as $\tilde{\alpha}=\alpha/N_1$; iii) Different from Table-based update in Algorithm~\ref{Algo:TD},  function approximation may result in very close estimated state values for adjacent states. This may result in cyclic path with the $\epsilon$-greedy action \eqref{eq:epsGreedy} between adjacent states like $\mathbf s\rightarrow \mathbf s^\prime \rightarrow \mathbf s$, which is obviously undesired. A simple remedy to this is to keep a copy of the previous state $\mathbf s_{\mathrm{prev}}$, and \eqref{eq:epsGreedy} is slightly revised by excluding the action that would lead $\mathbf s^\prime$ to $\mathbf s_{\mathrm{prev}}$; iv) Similar to Algorithm~\ref{Algo:TD}, the parameter $\boldsymbol \theta$ should be initialized so as to encourage the shortest-path flying at the first episode. To this end, $\boldsymbol \theta$ is initialized to the least square solution by minimizing $\|-\mathbf d(\tilde{\mathcal S})-\mathbf X^T(\tilde{\mathcal S})\boldsymbol \theta\|^2$, where $\tilde{\mathcal S}\subset \mathcal S$ is a selected subset of the state space to initialize $\boldsymbol \theta$, $\mathbf X(\tilde{\mathcal S})$ is the matrix with the feature vectors $\mathbf x(\mathbf s)$, $\mathbf s\in \tilde{\mathcal S}$ as the columns, and $\mathbf d(\tilde{\mathcal S})$ is the vector with the distances to $\mathbf q_F$ as the elements.

\section{Numerical Results}\label{sec:Numerical}
\begin{figure}
\centering
\includegraphics[width=0.3\textwidth]{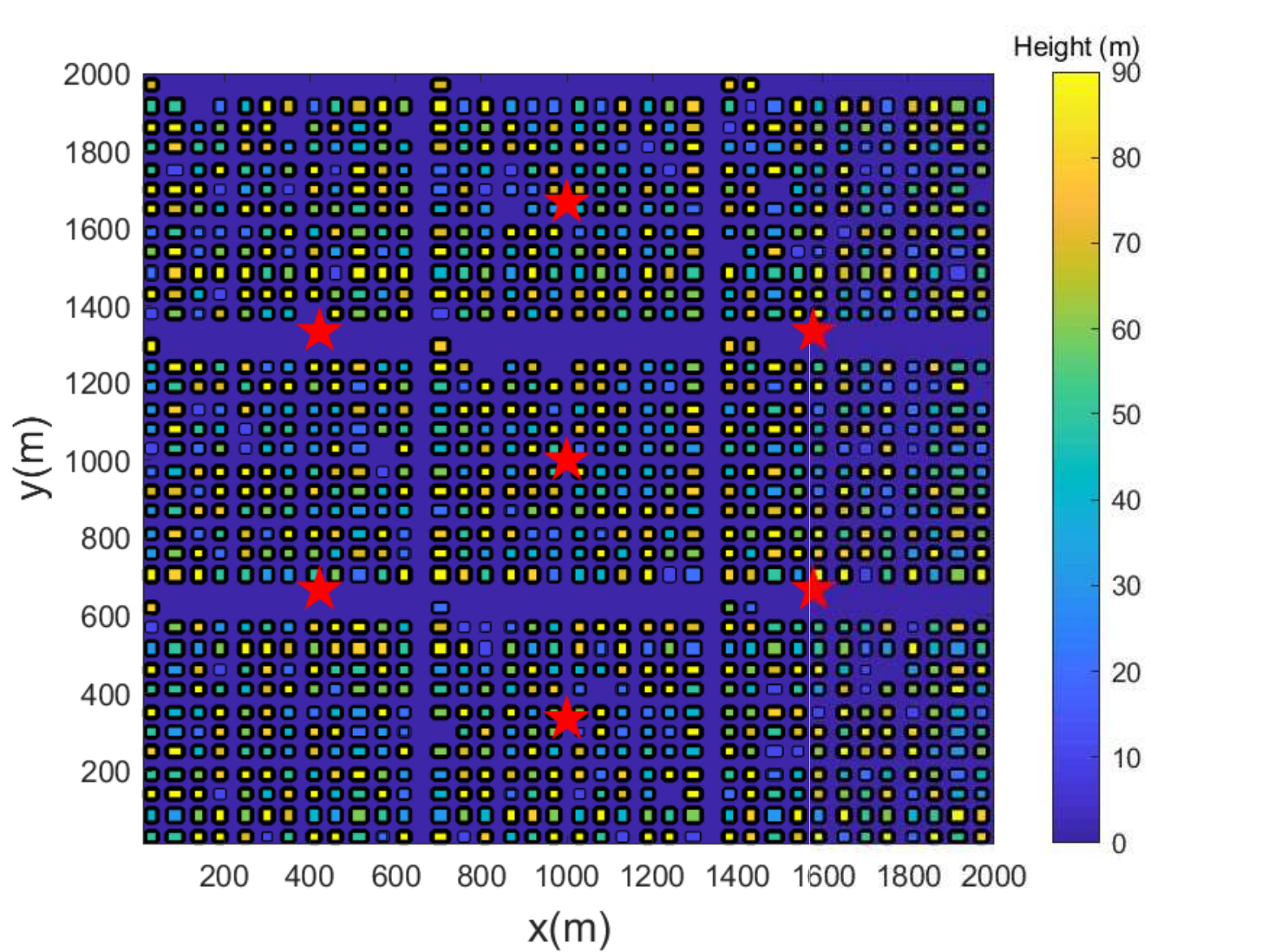}
\caption{The building locations and heights.\vspace{-2ex}}
\label{F:building}
\end{figure}

Numerical results are provided to evaluate the performance of the proposed UAV path designs. As shown in Fig.~\ref{F:building}, we consider an urban area of size 2 km $\times$ 2 km with high-rise buildings, which constitute the most challenging environment for communication-aware UAV path design, since the LoS/NLoS links and the received signal strength may alter frequently as the UAV flies (see Fig.~\ref{F:SystemModel}). To accurately simulate the BS-UAV channels, we first generate the building locations and height based on {\it one realization} of the statistical model suggested by the International Telecommunication Union (ITU) \cite{1094}, which involves three parameters: $\alpha_{\mathrm {bd}}$: the ratio of land area covered by buildings to the total land area; $\beta_{\mathrm{bd}}$: the mean number of buildings per unit area; and a variable determining the building height distribution, which is usually modelled as Rayleigh with mean $\sigma_{\mathrm{bd}}$. 
 Fig.~\ref{F:building} shows the realization of the building locations and height with $\alpha_{\mathrm {bd}}=0.3$, $\beta_{\mathrm{bd}}=300$ buildings/km$^2$, and $\sigma_{\mathrm{bd}}=50$ m. For simplicity, all building height is clipped to below $90$m.

We assume a hexagonal cell layout with two tiers in the considered area, which corresponds to 7 BS sites with locations marked by red stars in Fig.~\ref{F:building}, and the BS antenna height is $25$ m \cite{1012}. With the standard sectorization technique, each BS site contains 3 sectors/cells. Thus, the total number of cells is $M=21$. The BS antenna model follows the 3GPP specification \cite{1024}, where an 8-element  uniform linear array (ULA) is placed vertically with pre-determined phase shift to electrically downtilt the main lobe by $10^\circ$. This leads to the directional antenna with fixed 3D  radiation pattern, which is shown in Fig.~4 of \cite{1095}. To obtain the average signals received by the UAV from each cell, at each possible UAV location, we firstly determine whether there exists a LoS link between the UAV and the BS according to the building information, and then use the 3GPP BS-UAV path loss model for urban Macro (UMa) given in Table B-2 of \cite{1012}. 

%

\begin{figure}
\centering
\includegraphics[width=0.35\textwidth]{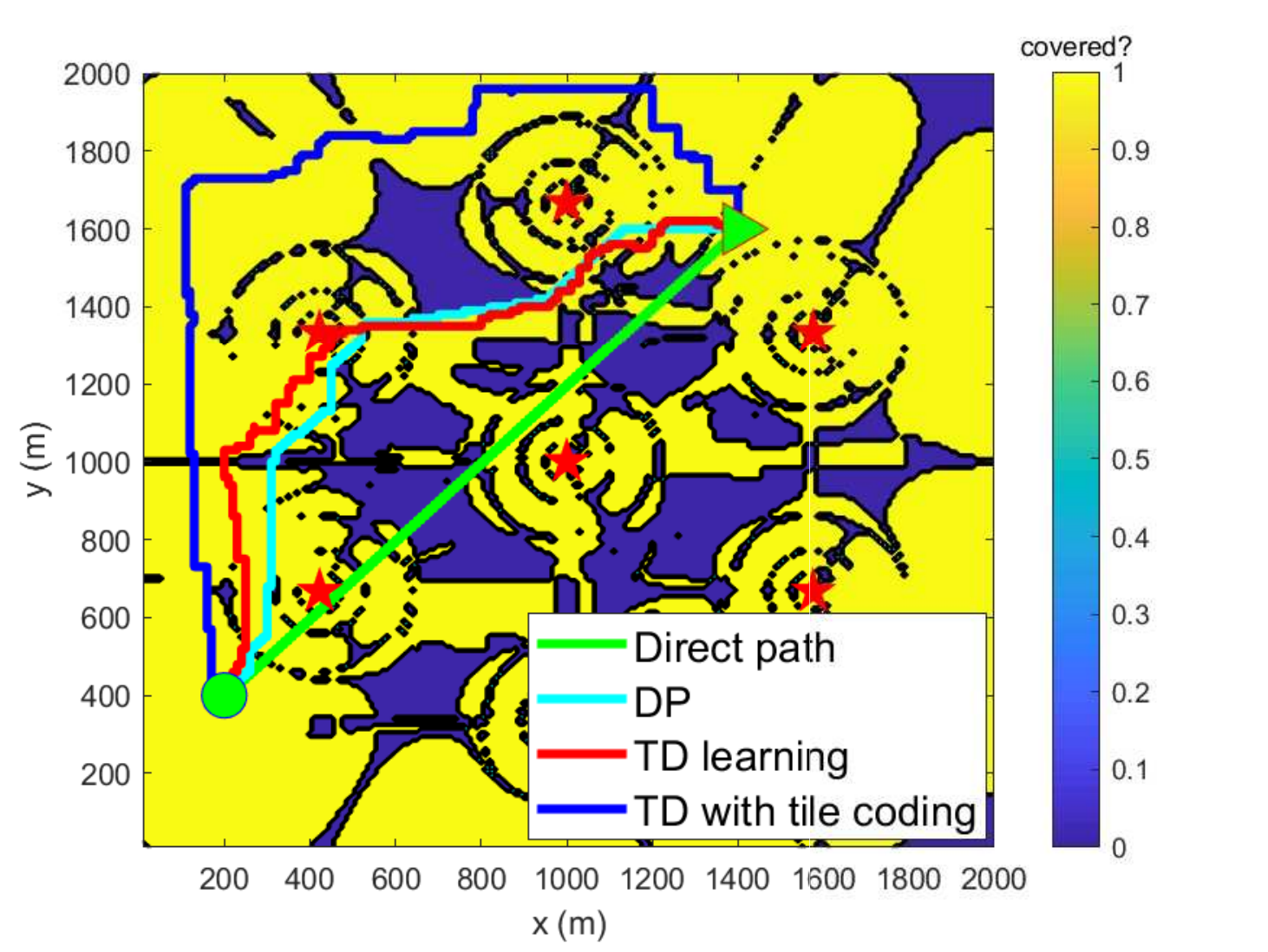}
\caption{The global coverage map and the resulting UAV paths.}
\label{F:signalMap}
\end{figure}

We assume that the UAV's flying altitude is $H=100$ m, and the SIR defined in Section~\ref{sec:SystemModel} is used as the performance measure to determine the cellular connectivity by the UAV. Fig.~\ref{F:signalMap} shows the global coverage map with $P_m=20$ dBm and $\gamma_{\thr}=0$ dB,  together with the resulting UAV paths from the initial location $\mathbf q_I=[200, 400]^T$ m to the final location $\mathbf q_F=[1400, 1600]^T$ m with four schemes: i) the direct path from $\mathbf q_I$ to $\mathbf q_F$; ii) the value-iteration based DP, which requires the perfect global coverage map; iii) the TD learning method proposed in Section~\ref{sec:TD}; and iv) TD learning with tile coding proposed in Section~\ref{sec:TDTileCoding}. The following parameters are used: $\mu=30$,  $K=4$, $\Delta=10$ m, $N_\alpha=2000$, $N_\epsilon=300$,  $\bar{N}_{\mathrm{epi}}=6000$ and $\bar N_{\mathrm{step}}=1000$. For tile coding, the number of tilings is $N_1=20$, and each tile has size $X_0 \times Y_0=200$ m $\times 200$ m. It is observed from Fig.~\ref{F:signalMap} that except the benchmark direct flight, the other three schemes all successfully find UAV paths that avoid the coverage holes of the cellular network. Furthermore, the table-based TD learning scheme gives a similar path as the optimal DP scheme. It is also noted that for TD with tile coding, a more conservative path with longer flying distance is obtained, since with linear function approximation, it seems more challenging to discover the narrow ``bridge'' as taken by the other two methods.

\begin{figure}
\centering
\includegraphics[width=0.25\textwidth]{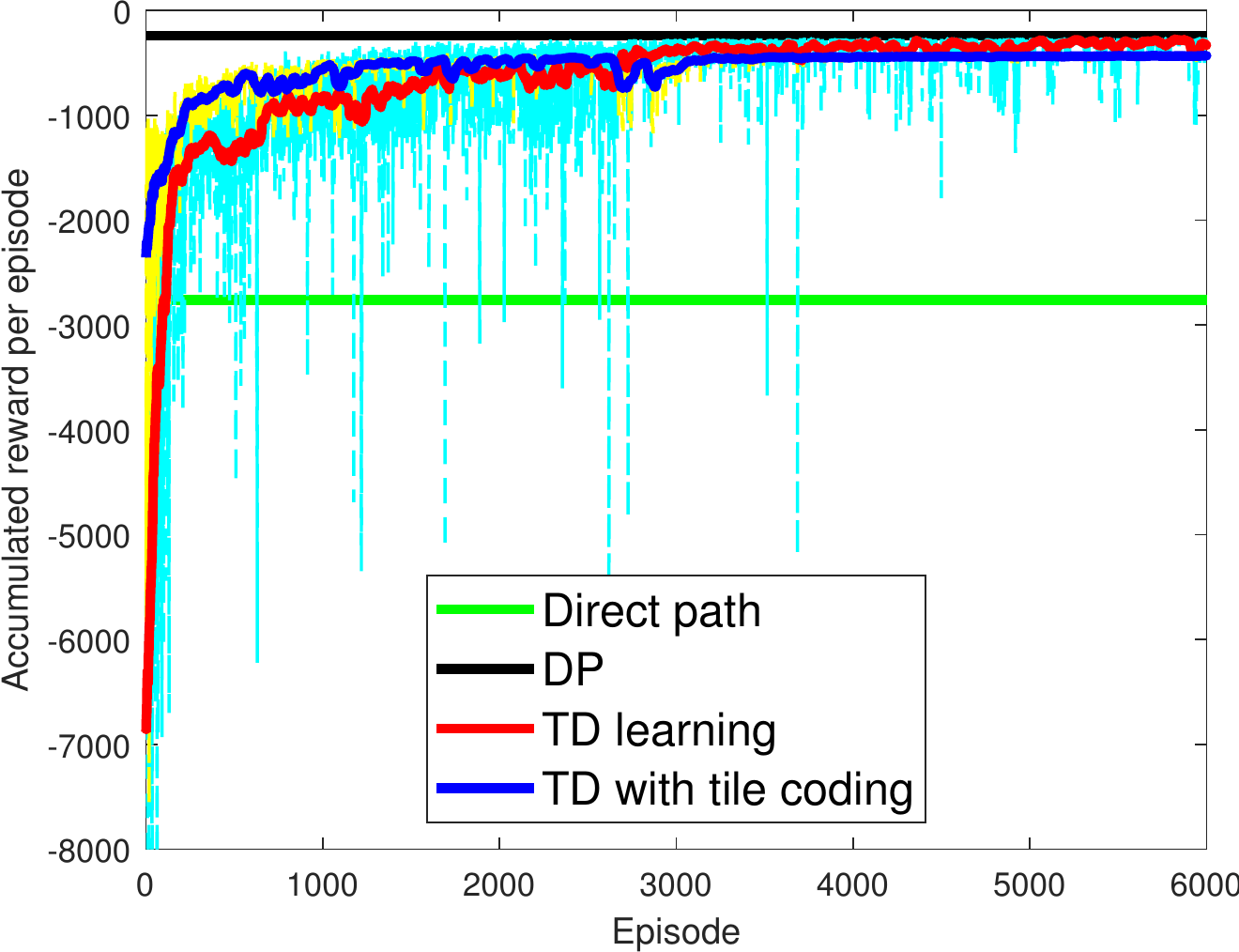}
\caption{Accumulated rewards per episode.\vspace{-3ex}}
\label{F:AccumulatedRewards}
\end{figure}

Fig.~\ref{F:AccumulatedRewards} shows the accumulated reward per episode for the TD learning algorithms. It is observed that both TD learning methods converge to values very close to the optimal DP solution, which significantly outperform the benchmark direct flight. It is also observed that tile coding helps improve the convergence speed of the TD learning method, though it eventually gives slightly worse performance. Lastly, it is observed that both TD learning methods require thousands of episodes to converge. 
 This gives rise to the typical issue of RL, i.e., learning from real experience is usually sample expensive. Fortunately, such issue can be alleviated by firstly pre-training the policy with simulation-generated samples according to certain (even inaccurate) communication model, which is almost cost-free, and then further refine the policy by actual UAV flight with online learning to address the model inaccuracy issue. 

\section{Conclusions}
This paper studies path designs for cellular-connected UAVs. 
 To overcome the limitations of conventional optimization-based path design approaches, we propose RL-based algorithms, which  only require  the measured or simulation-generated raw signal  strength as the input and are suitable for both online and offline implementations. The proposed algorithm utilizes the TD  method to learn the  state-value function, and it is further extended by applying linear function approximation with tile coding. Numerical results are provided to show the effectiveness of the proposed algorithms.

\bibliographystyle{IEEEtran}
\bibliography{IEEEabrv,IEEEfull}

\end{document}